\documentclass[fleqn,usenatbib]{mnras}
\usepackage{graphicx}
\usepackage[T1]{fontenc}
\usepackage{ae,aecompl}

\usepackage{amsmath}	
\usepackage{amssymb}	

\usepackage{newtxtext,newtxmath}
\usepackage{epstopdf}

\usepackage{url}

\hypersetup{draft}

\usepackage{upgreek}
\usepackage[usenames]{color} 
\usepackage{appendix}
\usepackage{color}
\usepackage{float}

\title[HD~42477: coupled modes]{HD~42477: coupled r~modes, g~modes and a p~mode in an A0Vnne star}

\author[D.W. Kurtz et al.]{D. W. Kurtz$^{1,2}$\thanks{E-mail: kurtzdw@gmail.com}, R. Jayaraman$^3$, P. Sowicka$^4$, G. Handler$^{4}$,  H. Saio$^5$,  \newauthor J. Labadie-Bartz$^{6,7,8}$,  U. Lee$^5$
\\
$^{1}$Department of Physics, North-West University, Dr Albert Luthuli Drive, Mahikeng 2735, South Africa\\
$^{2}$Jeremiah Horrocks Institute, University of Central Lancashire, Preston PR1 2HE, UK\\
$^3$Department of Physics, and Kavli Institute for Astrophysics and Space Research, M.I.T., Cambridge, MA 02139, USA\\
$^{4}$Nicolaus Copernicus Astronomical Center, Polish Academy of Sciences, ul. Bartycka 18, 00-716, Warszawa, Poland\\
$^{5}$Astronomical Institute, Graduate School of Science, Tohoku University, Sendai 980-8578, Japan\\
$^{6}$LESIA, Paris Observatory, PSL University, CNRS, Sorbonne University, Universit\'e Paris Cit\'e, 5 place Jules Janssen, 92195 Meudon, France\\
$^{7}$Homer L. Dodge Department of Physics and Astronomy, University of Oklahoma, 440 W. Brooks Street, Norman, OK 73019, USA\\
$^{8}$Instituto de Astronomia, Geof\'isica e Ci\^encias Atmosf\'ericas, Universidade de S\~ao Paulo, Rua do Mat\~ao 1226, \\Cidade Universit\'aria, 05508-900 S\~ao Paulo, SP, Brazil\\
}

\date{Accepted XXX. Received YYY; in original form ZZZ}
\pubyear{2023}

\begin{document}
\label{firstpage}
\pagerange{\pageref{firstpage}--\pageref{lastpage}} 
\maketitle

\begin{abstract}
Several studies have shown that a number of stars pulsating in p~modes lie between the $\beta$\,Cep and $\delta$~Sct instability strips in the Hertzsprung-Russell (HR) Diagram. At present, there is no certain understanding of how p~modes can be excited in this $T_{\rm eff}$ range. The goal of this work is to disprove the conjecture that all stars pulsating in p~modes and lying in this $T_{\rm eff}$ range are the result of incorrect measurements of $T_{\rm eff}$, contamination, or the presence of unseen cooler companions lying in the $\delta$~Sct instability strip (given the high binary fraction of stars in this region of the HR~Diagram). Using TESS data, we show that the A0Vnne star HD~42477 has a single p~mode coupled to several r~modes and/or g~modes. We rule out a contaminating background star with a pixel-by-pixel examination, and we essentially rule out the possibility of a companion $\delta$~Sct star in a binary. We model the pulsations in HD~42477 and suggest that the g~modes are excited by overstable convective core modes. We also conjecture that the single p~mode is driven by coupling with the g~modes, or that the oblateness of this rapidly-rotating star permits driving by \ion{He}{ii} ionization in the equatorial region. 
\end{abstract}  

\begin{keywords}
stars: emission-line, Be; stars: oscillations; asteroseismology; stars: individual: (TIC~294125876; HD~42477)
\end{keywords}

\section{Introduction}

In asteroseismology, pulsating stars are grouped into classes based on a combination of their positions in the Hertzsprung-Russell (HR) Diagram and their pulsation characteristics (e.g., \citealt{2021RvMP...93a5001A,2022ARA&A..60...31K}; \citealt[chapter 2]{2010aste.book.....A}). The main groups along the upper main-sequence, for pulsating stars of A--F spectral type, are the g~mode pulsators ($\gamma$~Dor stars), and the p-mode and mixed p- and g-mode pulsators ($\delta$~Sct stars). The observed and theoretical instability strips for these two groups overlap. Among B stars, the main groups are the hotter  p-mode and mixed p- and g-mode pulsators ($\beta$\,Cep stars), and the cooler g~mode pulsators (Slowly Pulsating B, or SPB, stars). The observed and theoretical instability strips for these two groups also have a small overlap. Classical Be stars can be found mixed in with these two classes; these stars range in spectral type from late-O to early-A. These Be stars are the most rapid rotators found among main-sequence stars, with equatorial rotation velocities nearing breakup velocity. These emission-line stars possess circumstellar disks that result from mass ejection episodes. There may exist a connection between observed nonradial g~mode pulsation behaviour in these stars and these outbursts (see, e.g., Section 7 of \citealt{2015MNRAS.450.3015K}, and \citealt{2018pas8.conf...69B, 2018A&A...610A..70B, 2021MNRAS.508.2002R}). 

For some time, observations have shown that there exist stars in all the OBAF pulsation classes -- $\gamma$~Dor, $\delta$~Sct, SPB and $\beta$\,Cep -- that appear to lie outside of their respective theoretical instability strips. Suggestions have been made to explain the pulsation driving in these stars, but there remains no clear consensus on the underlying mechanism. One conjecture  suggests that such pulsations require no astrophysical explanation; rather, they can be explained away by an incorrect effective temperature ($T_{\rm eff}$) value, background contaminators, or undetected binary companions. This conjecture can be refuted by proving that even one star lying between the $\beta$\,Cep and $\delta$~Sct instability strips pulsates in p~modes. This important task is difficult and has not previously been attempted.

In this paper, we aim to show that HD\,42477, which lies between the $\beta$\,Cep and $\delta$\,Sct instability strips, pulsates in p~modes. In Section\,\ref{sec:background}, we cite the numerous observational and theoretical studies that highlight the problem of understanding pulsation driving in stars lying outside the theoretical instability strips. In Section\,\ref{sec:hd42477} HD\,42477 is introduced in detail. Section\,\ref{sec:fa1} presents a frequency analysis of the TESS data for HD~42477 and discusses the coupling of a high-frequency p~mode to low frequency g~ and r~modes. In Section\,\ref{sec:px-by-px} we present a pixel-by-pixel analysis of the field-of-view for HD~42477 to rule out any potential contamination from a background star or from a wide orbital companion. Some $\delta$~Sct stars are known to show g~modes, r~modes and p~modes in the same star \citep[e.g.,][]{2018MNRAS.476.3169B,2018MNRAS.474.2774S}, so we must work to rule out the possibility that the observed frequency spectrum originates in a contaminating star. This is a critical portion of our analysis that we discuss in detail, given the high precision of the data, typically with amplitude uncertainties in the several $\umu$mag range for brighter stars. We then conduct a careful analysis in Section\,\ref{sec:specsearch} of spectra we obtained of HD~42477 to search for any orbital companions.  Finally, in Section 7 we present a pulsation model for HD~42477 identifying the low frequencies with g~modes and r~modes and discuss various phenomena that could drive the p~mode, given that this is the first p-mode pulsator proven to lie between the $\beta$\,Cep and $\delta$~Sct instability strips.

\section{Background to the problem}
\label{sec:background}
\subsection{Stars lying outside instability strips}

In a study of B stars observed by the {\it CoRoT} mission, \citet{2009A&A...506..471D} suggested the existence of a new class of pulsating variables in the spectral type range between the SPB and $\delta$~Sct instability strips, i.e., late-B to early-A spectral types. They also noted that the highest-amplitude SPB stars show both nonlinearities, in the form of combination frequencies, and low frequencies with excess amplitude after the principal pulsation frequencies have been pre-whitened. Such behaviour indicates amplitude and/or frequency modulation as a consequence of limited lifetimes for these modes that may be caused by nonlinear resonant mode locking. Other possibilities are that the low frequencies in these stars, in addition to g\,modes, are also the result of closely spaced, unresolved, or only partially resolved r~mode frequencies (see, e.g., \citealt{2018MNRAS.474.2774S}) and stochastically driven modes \citep{2019NatAs...3..760B,2020A&A...640A..36B}.

To analyse A--F type stars, \citet{2018MNRAS.476.3169B} extracted 983 $\delta$~Sct stars with 4 yr of data from the 200\,000 stars observed by {\it Kepler}. They found a significant number of $\delta$~Sct stars pulsating in p\,modes whose temperatures are higher than the theoretical blue edge of the $\delta$~Sct instability strip, as calculated by \citet{2004A&A...414L..17D} and \citet{2005A&A...434.1055G}. That edge is often presented as the theoretical hot boundary for these stars, but this depends on the value of the mixing length chosen for the models \citep{2015LRSP...12....8H,2018MNRAS.476.3169B}. However, no choice of mixing length produces a blue border that encompasses all of the $\delta$~Sct stars, as shown in \citet{2019MNRAS.485.2380M}. 
In this work, they analysed 15\,000 {\it Kepler} A and F stars and discussed the pulsator fraction of $\delta$~Sct stars across the instability strip and up to $T_{\rm eff} = 10\,000$\,K. \citeauthor{2019MNRAS.485.2380M} found $\delta$~Sct stars hotter than the blue edge; to explain these, they argued statistically that these cannot be accounted for by uncertainties in $T_{\rm eff}$. However, they did not address whether all $\delta$~Sct stars hotter than the theoretical blue border may be binaries that include a cooler $\delta$~Sct star. Evidence supporting the binary hypothesis can be found in Table\,13 of \citet{2017ApJS..230...15M}, which shows that the majority of stars with masses of $\sim 2 - 3$\,M$_\odot$ have companions with orbital periods greater than 2\,d and mass ratios $q > 0.1$. More than 40~per~cent of those have $q > 0.3$. Both of these relative fractions increase with higher main-sequence masses and hotter main-sequence $T_{\rm eff}$. 

To study $\delta$\,Sct stars in binaries, \citet{2018MNRAS.474.4322M} used the phase modulation technique (PM) to identify 341 non-eclipsing binaries with $P_{\rm orb} > 20$\,d in a sample of 2224 A and F stars in the {\it Kepler} data. Approximately 7~per~cent of those systems exhibited pulsations in both stars, suggesting that there a significant fraction of $\delta$~Sct stars can be found in binary systems where both stars are $\delta$~Sct. By extension, there must exist primary stars hotter than the blue border with cooler $\delta$~Sct secondaries, which could explain $\delta$~Sct stars hotter than the theoretical blue border. \citeauthor{2018MNRAS.476.3169B} also found that g~mode pulsations exist in $\delta$~Sct stars at temperatures higher than the theoretical $\gamma$~Dor g-mode instability strip, confirming the discovery of \citet{2010ApJ...713L.192G} that there are ``practically no pure $\delta$~Sct or $\gamma$~Dor pulsators.'' However, their study (and almost all others) did not consider r~modes, and most studies have not carefully distinguished combination frequencies due to mode coupling from g~mode frequencies, although this problem is widely recognised and discussed.
 
These observations of the B and A stars with {\it CoRoT} and {\it Kepler} give rise to important questions. The primary driving mechanism across the upper main-sequence stars is the $\kappa$-mechanism -- in the \ion{He}{ii} ionization zone for $\delta$~Sct stars, and the in the ionization zone of Fe-peak elements for B stars. In addition, convective blocking is thought to be the dominant driving mechanism in $\gamma$~Dor stars \citep{2005A&A...435..927D}. However, if both p\,modes and g\,modes are found in stars outside of the theoretical instability strips, then we must further improve our understanding of driving and damping in main-sequence B, A and F stars. Suggesting that existing pulsation models must be modified is a bold claim, and there is a significant burden of proof required to show that the observations demand this.

\subsection{The advent of large-scale sky surveys}

To find and analyse further examples of stars that lie outside instability strips, we can utilize the data from large-scale sky survey missions such as {\it Kepler} and TESS, which have observed large regions of the sky and provided many months, and even years, of data for millions of stars. The larger TESS field-of-view and its observations of the galactic plane have allowed for the observation of more B and A stars than {\it Kepler}. Additionally, its focus on relatively bright stars has enabled extensive ground-based follow-up to determine accurate stellar parameters -- T$_{\rm eff}$, surface gravity ($\log\,g$), distance, luminosity and metallicity -- for these targets.

\citet{2020MNRAS.493.5871B} carried out a large-scale analysis of about 50\,000 stars observed by TESS and found a large number of p~mode pulsators in the temperature range between the $\beta$\,Cep and $\delta$~Sct instability strip. They labeled these stars ``Maia'' variables.\footnote{This name arises from historical reasons, but the prototype star Maia itself does not pulsate. This calls for a different name for such stars.} \citeauthor{2020MNRAS.493.5871B} recognised two problems with making conclusions from the apparent positions in the HR~Diagram of this suggested new class of pulsating variable stars: There could be errors in the positions caused by uncertainties in $T_{\rm eff}$, and the stars may be in binaries. They explicitly mention this latter possibility and suggest that Maia variables may consist of a non-pulsating B star and a $\delta$~Sct star, but they do not test individual stars in their sample. Another potential issue they neglect to mention is the presence of background contaminating stars, which often cause problems due to TESS's large plate scale ($21\arcsec$/pixel). Our work discusses these three issues in detail.

\subsection{Possible explanations for these stars}

Both \citet{2018MNRAS.476.3169B} and \citet{2020MNRAS.493.5871B} used spectroscopic temperatures to claim that there exist p~mode pulsators in the range between the $\beta$\,Cep and $\delta$~Sct stars. They argued that the uncertainties in $T_{\rm eff}$ are sufficiently low such that the stars are not $\beta$\,Cep or $\delta$~Sct pulsators that are incorrectly plotted in the HR~Diagram due to a poorly estimated $T_{\rm eff}$. Similarly, as mentioned above, \citet{2019MNRAS.485.2380M} made statistical arguments that errors in $T_{\rm eff}$ cannot explain $\delta$~Sct stars that are hotter than the theoretical blue border. These three studies have ruled out the explanation that an incorrect T$_{\rm eff}$ is the source of p~mode pulsators found to lie outside the instability strips for $\beta$\,Cep and $\delta$\,Sct stars.

Could binarity explain these p~mode pulsators? This question has not been well-examined; however, it has been shown that the binary fraction for B stars is essentially 100~per~cent \citep{2017ApJS..230...15M}, and many companions to B stars could be A stars in the $\delta$~Sct instability strip. Perhaps the p~mode pulsators found outside the instability strips could all have unrecognised $\delta$~Sct companions in orbits ranging from tight to wide. While it is difficult to resolve this using TESS or previous observations, we can disprove this explanation by showing that at least one star in the correct T$_{\rm eff}$ range between those two instability strips pulsates with at least one p~mode. Such proof involves the observationally difficult task of disproving binarity or background contamination.

While incorrect parameters, unresolved binarity, or observational issues can result in stars lying outside the classical instability strips, the possibility that the instability strip boundaries should be extended must also be seriously considered. \citet{2017MNRAS.469...13S} recalculated the instability strips of upper main-sequence stars with masses from $2 - 20$\,M$_\odot$. They examined modes of degree $\ell \le 4$ with the effects of rotation included, which shifts the boundary of the SPB instability strip to a lower temperature. They conjectured that this, combined with surprisingly high frequencies for rapid rotators, may explain the p\,mode pulsators that appear to lie between the SPB and $\delta$~Sct instability strips. \citet{2014A&A...569A..18S} also conjectured that gravity darkening in rapid rotators, which can make stars appear cooler when viewed equator-on, as well as higher frequencies from sectoral prograde dipole modes, can explain apparent g\,mode pulsators that are observed in the SPB -- $\delta$~Sct gap in NGC\,3766 \citep{2013A&A...554A.108M}.

Similar to the stars lying between the $\beta$\,Cep and $\delta$\,Sct instability strips, there also exist hot $\gamma$~Dor stars with temperatures well above the theoretical hot border of that instability strip. \citet{2020MNRAS.493.4518K} studied 24 such stars spectroscopically, testing whether they could be explained as binary stars with a hotter $\delta$~Sct and cooler $\gamma$~Dor star, or whether they could be rapidly rotating SPB pulsators with a lower T$_{\rm eff}$ due to gravity darkening. They concluded that the number of hot $\gamma$~Dor stars is low, but real; due to the shallowness of their surface convective zones (implying that convective blocking cannot drive pulsations), there must exist an additional driving mechanism in some A stars with g\,mode pulsations. Indeed, in some $\gamma$\,Dor stars, the pulsation modes are a combination of g~modes and r~modes -- global Rossby waves \citep{2016A&A...593A.120V}. \citet{2018MNRAS.474.2774S} found r~modes in B and A stars, including an outbursting Be star; \citet{2020MNRAS.497.4117L} then showed how a rotating convective core can couple to g~modes in the radiative envelope of a 2-M$_\odot$ model, thus explaining the commonly-observed low frequencies seen in the amplitude spectra of B and A stars that lie close to their respective inferred rotation frequencies. 

Our work describes a 9900\,K model of HD~42477 explaining the low frequencies as g~modes and r~modes; we suggest that the latter are a common feature in Be stars. We also show that the low frequency r~modes and g~modes in this star are nonlinearly coupled to the high frequency p~mode, which conclusively proves that the p~mode is in the same star as the g~modes and r~modes. This coupling of modes (for which there exists a theoretical understanding) may provide the boost in driving needed to understand how p~modes can be excited in the T$_{\rm eff}$ range between the $\beta$\,Cep and $\delta$~Sct theoretical instability strips. Previous calculations of the boundaries of those strips have not considered such a boost from coupled modes. 

\section{HD~42477 and its Observations}
\label{sec:hd42477}

\begin{figure*}
\begin{center}
\includegraphics[width=0.9\linewidth,angle=0]{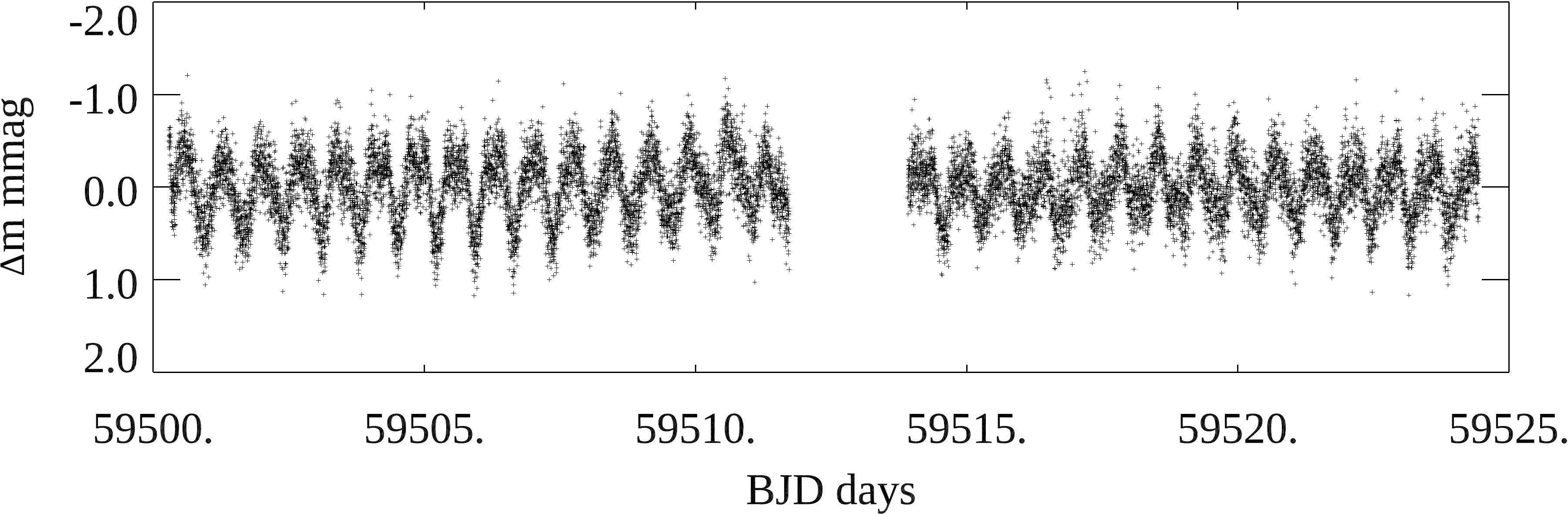}
\caption{The light curve of HD~42477 obtained by TESS in 2-min cadence in Sector 44. The ordinate scale is Barycentric Julian Date -- 2\,400\,000.0. The low frequency variations are clear. The other Sectors look similar; only this one is shown for reasons of scale.}
\label{fig:lc} 
\end{center}
\end{figure*}

HD~42477 (TIC~294125876, HR~2191) is classified as A0Vnne in the work of \citet{1969AJ.....74..375C} due to its broad spectral lines (implying rapid rotation) and the presence of emission lines. Hence, this is an archetypal Be star at the cool end of the range. Indeed, \citeauthor{1969AJ.....74..375C} note a ``sharp K line superimposed on [a] hazy one'', which is consistent with a Be circumstellar disk; they further note ``Si:'', which signifies marginal enhancement of Si. It is unlikely that this star is a marginal Ap star with weak abundance anomalies, as both magnetic (ApSi) and non-magnetic, or weakly magnetic, (HgMn) Ap stars rotate much more slowly than HD~42477 and generally show rotational light variations from surface spots \citep{2005A&A...439.1093K}, which are not present for HD~42477. 

Archival spectroscopy of HD~42477 from the Be Star Spectra (BeSS) database\footnote{http://basebe.obspm.fr/basebe/} \citep{Neiner2011} taken between 2007 and 2022 show that the disk was already weak in 2007 (with the double-peaked emission in H$\alpha$ reaching only up to $\sim$0.85 in continuum units within the broader absorption profile), and that it had been steadily dissipating through at least the end of the TESS observations considered here.
To corroborate this,  we obtained 12 high-resolution (R$\sim$53000), high signal-to-noise ratio ($\sim 300 - 400$; $t_{\rm exp} = 1800$\,s) echelle spectra between 2021 May 07 -- 2021 Dec 01 (which overlapped with TESS sectors 43-45) using the NRES instrument attached to the 1-m telescopes at the Wise, CTIO and McDonald observatories operated by the Las Cumbres Observatory network \citep{2013PASP..125.1031B}. Analysis of these data (also discussed in Section \ref{sec:specsearch}) indicates that there was no circumstellar activity related to ongoing stellar mass ejection that might otherwise contaminate the variations seen by TESS, so that all photometric variability is stellar in origin, i.e., pulsational. Further, such a weak disk (as is typical for the least massive Be stars) should produce virtually no excess continuum flux at visible wavelengths \citep{Vieira2017} so that the derivation of the stellar parameters in Sec. 7.1 should be wholly unaffected by the disk.

The most consequential aspect of HD 42477 being a classical Be star lies in the pulsational properties of Be stars at large. Space photometry has revealed that virtually all Be stars pulsate \citep{Rivinius2016, Semaan2018, Labadie-Bartz2022}. From analyses of 432 TESS Be stars, \citeauthor{Labadie-Bartz2022} showed that almost all Be stars exhibit multiple peaks in amplitude spectra in the $0.5 - 4$\,d$^{-1}$ range, the usual range for g~modes. Closely spaced groups of frequencies were present in over 85~per~cent of their 432 stars; \citet{2015MNRAS.450.3015K} showed that these frequency groups often are combination frequencies resulting from nonlinear coupling of pulsation modes. Thus, multi-mode, non-radial pulsation is common in Be stars.

To analyse the pulsations of HD~42477, we used 2-min cadence TESS data obtained during Sectors 33 and 43-45. The data are available in both SAP (simple aperture photometry) and {\mbox PDCSAP} (presearch-data conditioning SAP); we used the S33 {\mbox PDCSAP} data for modelling, and the S43-45 {\mbox PDCSAP} data for frequency analysis, after converting intensity to magnitudes. The S43-45 data have a time span of 75.46\,d and comprise 48060 data points (after clipping 28 outliers), with a centre point in time of $t_0 = {\rm BJD}~24559512.40014$. The $\sim$2-d gaps in the data at the time of perigee correspond to the period of data downlink, during which no observations were taken. Fig.\,\ref{fig:lc} shows the light curve of the Sector 44 data, which demonstrates clear low-frequency variability. 

\section{Frequency analysis of HD~42477}
\label{sec:fa1}

\begin{figure*}
\begin{center}
\includegraphics[width=0.9\linewidth,angle=0]{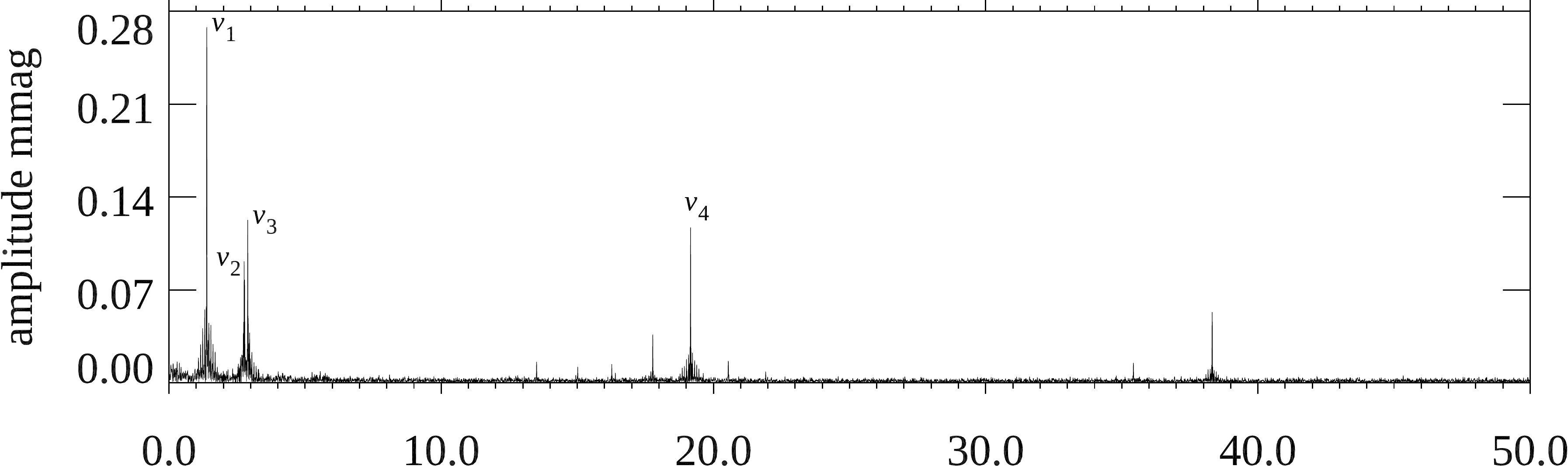}
\includegraphics[width=0.9\linewidth,angle=0]{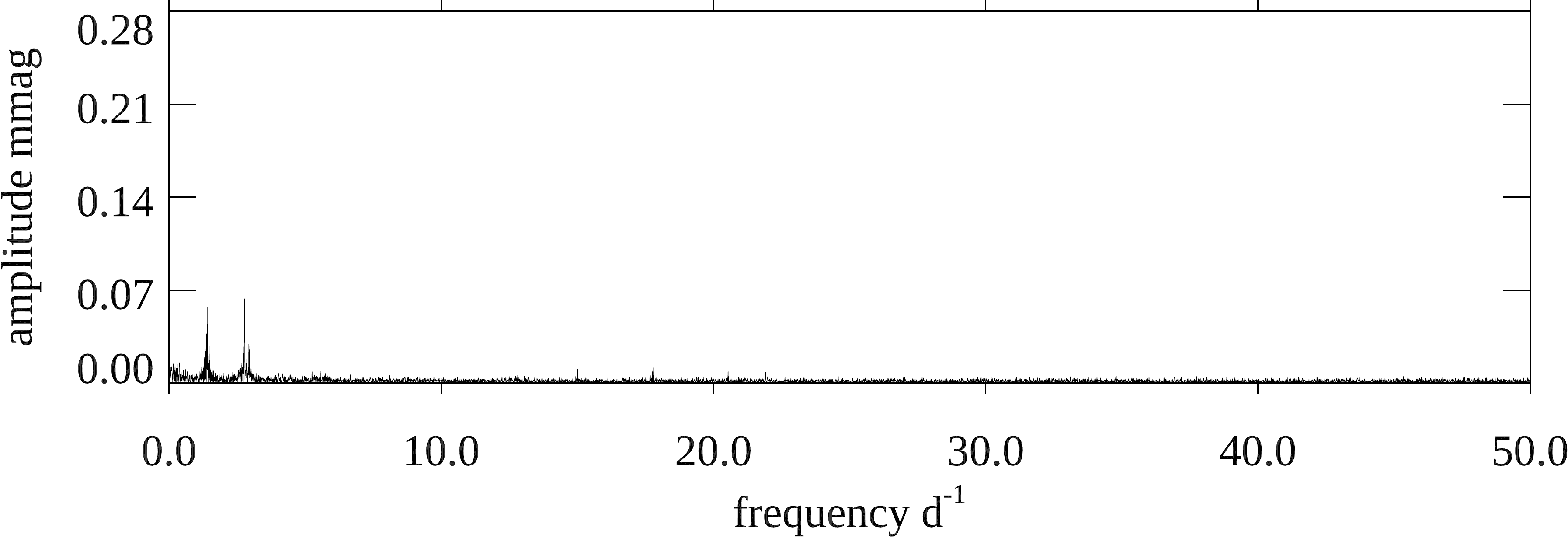}
\caption{Top: An amplitude spectrum of the Sector 43-45 data for HD~42477, showing the four identified mode frequencies, other combination frequencies, and a harmonic in the p~mode frequency range. Bottom: an amplitude spectrum of the residuals after pre-whitening the 4 mode frequencies, plus the harmonic of $\nu_4$, and their exact combinations. The residual amplitude at low frequency is caused by amplitude modulation of the low frequencies, which contributes to some modulation of the combination frequencies, resulting in some residual amplitude for those as well. }
\label{fig:ft1} 
\end{center}
\end{figure*}

\begin{figure}
\begin{center}
\includegraphics[width=0.75\linewidth,angle=0]{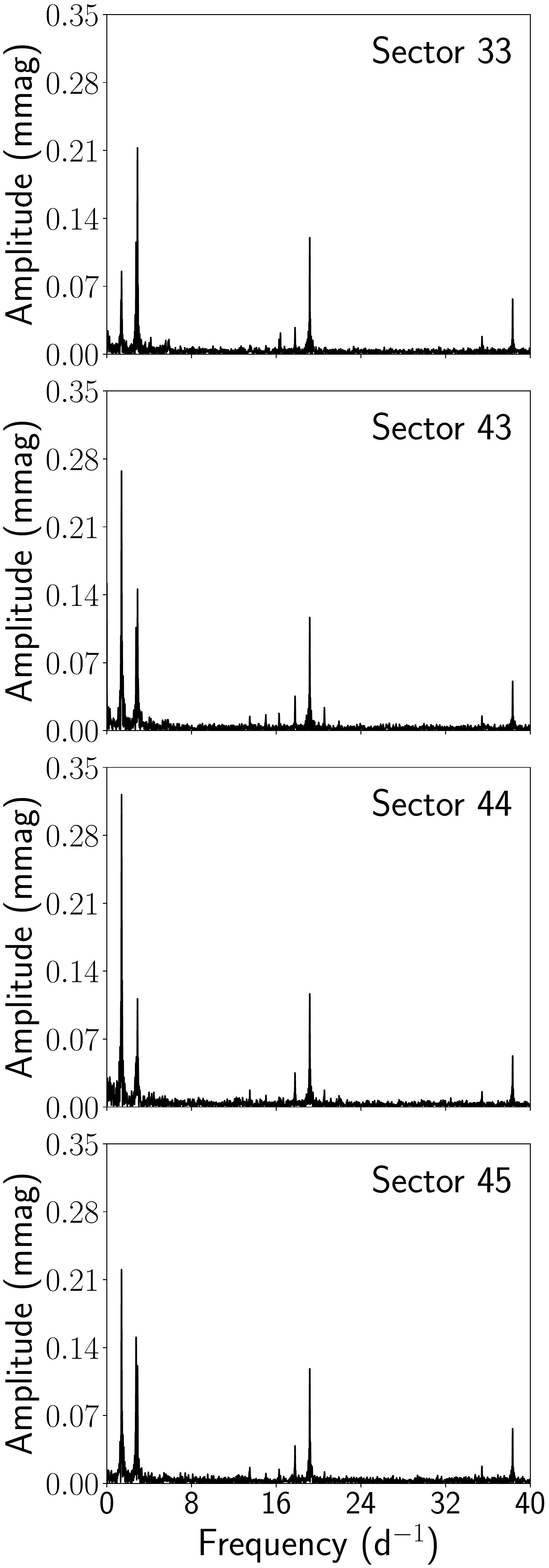}
\caption{Amplitude spectra of the Sector 33 and 43-45 data for HD~42477. The amplitude variations in the low frequencies, $\nu_1$, $\nu_2$ and $\nu_3$, is evident -- especially the month-to-month variability between Sectors 43, 44 and 45. The amplitude of the p~mode, $\nu_4$, remains stable throughout.}
\label{fig:ft2} 
\end{center}
\end{figure}

The S43-45 data were analysed using a fast Discrete Fourier Transform (see, e.g., \citealt{1985MNRAS.213..773K}) to produce the amplitude spectra shown in Fig.\,\ref{fig:ft1}. Thirteen significant resolved peaks (listed in Table \ref{Tab:nld}) were identified. The 3 low-frequency peaks in the $1-3$\,d$^{-1}$ range ($\nu_1$, $\nu_2$, $\nu_3$) do not have any harmonic relationship. There are 10 frequencies in the $\delta$~Sct pulsation range between $13-40$\,d$^{-1}$. Table 1 demonstrates that there is one p~mode pulsation frequency, labelled $\nu_4$, and there exist 8 other frequencies in the p~mode range that are consistent with combinations of $\nu_4$ with the 3 resolved low-frequency peaks -- $\nu_1$, $\nu_2$ and $\nu_3$. One other frequency listed in the table is a harmonic of $\nu_4$. Importantly, the presence of combination frequencies between the p~mode and the low frequency peaks demonstrates that the $\delta$~Sct p~mode pulsation is in the same star that produces the low frequency peaks. Moreover, the low-frequency signals in HD 42477 are completely typical for a Be star, supporting the hypothesis that they originate in the rapidly-rotating A0Vnne star.

\begin{table*}
\centering
\caption{A non-linear least squares fit of the frequencies derived from S43-45 for HD~42477. The zero point for the phases is $t_0 = {\rm BJD}~24559512.40014$. Note that there is only one independent p~mode pulsation frequency at $\nu_4 =  19.16001$\,d$^{-1}$.  The last column shows the difference between the derived frequency and an exactly calculated frequency from the combination identifications in column 1, divided by the Rayleigh frequency resolution for the S43-45 data set of $0.013$\,d$^{-1}$, thus demonstrating the frequency matches to the identifications.}
\begin{tabular}{lrrrr}
\hline
\hline
&\multicolumn{1}{c}{frequency} & \multicolumn{1}{c}{amplitude} &   
\multicolumn{1}{c}{phase}  & $\Delta \nu / \sigma$\\
&\multicolumn{1}{c}{d$^{-1}$} & \multicolumn{1}{c}{mmag} &   
\multicolumn{1}{c}{radians}  & \\
& & \multicolumn{1}{c}{$\pm 0.0014$} &   &
   \\
\hline
$\nu_1$ & $1.39026 \pm 0.00004 $ & $0.2678 $ & $-0.213 \pm 0.005 $ & $ $ \\
$\nu_2$ & $2.76160 \pm 0.00011 $ & $0.0884 $ & $-2.169 \pm 0.016 $ & $ $ \\
$\nu_3$ & $2.89400 \pm 0.00009 $ & $0.1191 $ & $-1.518 \pm 0.012 $ & $ $ \\
$\nu_4 - 2\nu_3 - \nu_2$ & $13.50479 \pm 0.00068 $ & $0.0159 $ & $-1.080 \pm 0.097 $ & $0.0 $ \\
$\nu_4 - \nu_2 - \nu_1$ & $15.01599 \pm 0.00081 $ & $0.0124 $ & $-2.783 \pm 0.115 $ & $-0.6 $ \\
$\nu_4 - \nu_3$ & $16.26642 \pm 0.00075 $ & $0.0135 $ & $-2.644 \pm 0.106 $ & $0.0 $ \\
$\nu_4 -\nu_2$ & $16.39813 \pm 0.00147 $ & $0.0069 $ & $2.906 \pm 0.207 $ & $0.0 $ \\
$\nu_4 -\nu_1$ & $17.77253 \pm 0.00028 $ & $0.0361 $ & $1.063 \pm 0.040 $ & $-0.2 $ \\
$\nu_4$ & $19.16001 \pm 0.00009 $ & $0.1166 $ & $0.164 \pm 0.012 $ & $ $ \\
$\nu_4 + \nu_1$ & $20.54661 \pm 0.00065 $ & $0.0156 $ & $-1.609 \pm 0.091 $ & $0.3 $ \\
$\nu_4 + 2\nu_1$ & $21.91939 \pm 0.00123 $ & $0.0082 $ & $1.978 \pm 0.174 $ & $1.6 $ \\
$2\nu_4 - \nu_3$ & $35.42660 \pm 0.00068 $ & $0.0147 $ & $0.082 \pm 0.097 $ & $0.0 $ \\
$2\nu_4$ & $38.32018 \pm 0.00019 $ & $0.0534 $ & $2.941 \pm 0.027 $ & $0.0 $ \\
\hline\hline
\end{tabular}
\label{Tab:nld}
\end{table*}

\subsection{The low frequency g~modes and r~modes}
\label{sec:lowf}

Because the low frequencies are not harmonically related, we rule out both binary and spot light curves as the source of these peaks in the amplitude spectrum, as well as for the variability seen by eye in Fig.\,\ref{fig:lc}. In a combined analysis of all the data from Sectors 43-45, pre-whitening all the frequencies presented in Table~1 yields some residual amplitude at these frequencies, as can be seen in the bottom panel of Fig.\,\ref{fig:ft1}. This indicates amplitude modulation of the r~modes or g~modes, or beating in unresolved modes on a timescale of months. This amplitude modulation is clearly visible in Fig.\,\ref{fig:ft2}. However, the single p~mode pulsation peak and its exactly-calculated combinations and harmonic in the $\delta$~Sct frequency range are all removed in the pre-whitening, as shown in the bottom panel of Fig.\,\ref{fig:ft1}. The small residual peaks discernible in the p~mode range after pre-whitening are caused by amplitude modulation of the coupled modes. 

The behaviour of the low-frequency modes is unusual in this star.
The upper main-sequence stars presented in \citet{2018MNRAS.474.2774S} show a dense hump of r~mode peaks that are unresolved, whereas HD~42477 has only a few low frequency peaks. To examine the hypothesis that these observed low frequencies in HD~42477 result from both g~and r~modes, we modelled the pulsations (discussed in Section\,\ref{sec:model}).

\section{Pixel-by-pixel examination}
\label{sec:px-by-px}

Even though we have shown that all the modes arise in the same star, we must demonstrate that the pulsations observed in the TESS data do not arise from a background contaminator.

\begin{figure}
\begin{center}
\includegraphics[width=0.5\textwidth]{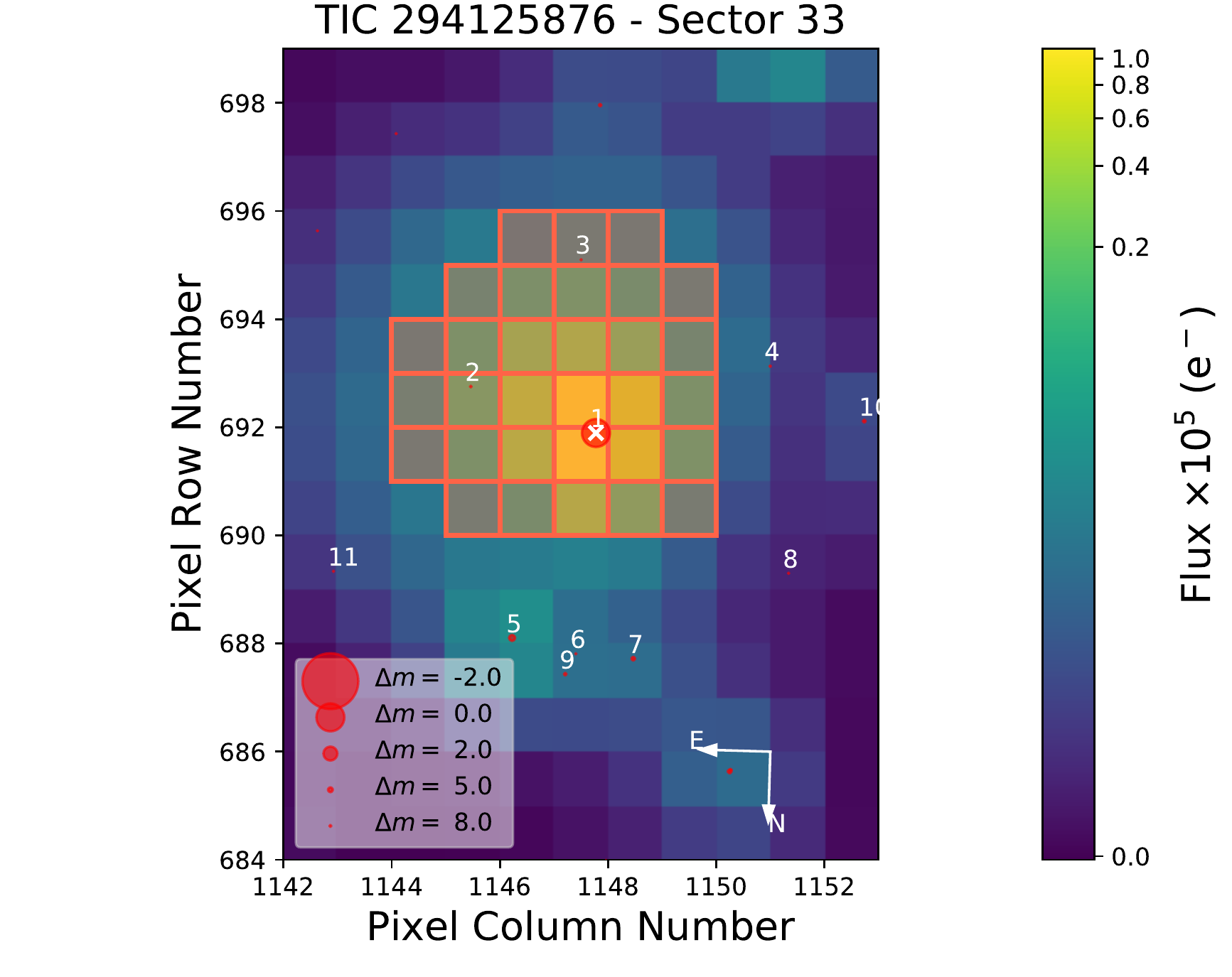} 
\caption{The TESS Target Pixel File of HD~42477 from Sector~33. All pixels within the SPOC pipeline mask are delineated by red squares. Gaia DR2 sources in the field are marked by red circles, where the size of the circle is a proxy for the Gaia magnitude; specifically, it represents the magnitude differential between HD\,42477 and the Gaia source. This plot was created with \texttt{tpfplotter}  \citep{2020A&A...635A.128A}.}. 
\label{fig:tpfgaia}
\end{center}
\end{figure}

For TESS photometry of bright stars such as  HD~42477, the optimal apertures chosen by the Science Processing Operations Center (SPOC) PDC pipeline \citep{jenkins-spoc} consist of many pixels (each of spatial size $21\arcsec$), introducing the risk of contamination from other objects located nearby on the sky. We downloaded the Target Pixel File (TPF) of HD~42477 from the Barbara A. Mikulski Archive for Space Telescopes (MAST)\footnote{\url{http://archive.stsci.edu}} and examined it using the Python package \texttt{lightkurve} \citep{2018ascl.soft12013L}. Fig.\,\ref{fig:tpfgaia} shows the TPF plot, with nearby Gaia targets up to 8 magnitudes fainter overplotted. We can see that all objects within this field are much fainter than our target and should not be significant sources of contamination. This is also reflected in the \texttt{CROWDSAP} keyword value, which characterises the fraction of total flux in the aperture that is estimated to arise from the target; this parameter has a value $> 0.997$ for all four sectors of data. Such a high value for this parameter suggests that any flux contribution from other sources is negligible. 

\begin{figure*}
\begin{center}
\includegraphics[width=1.0\textwidth]{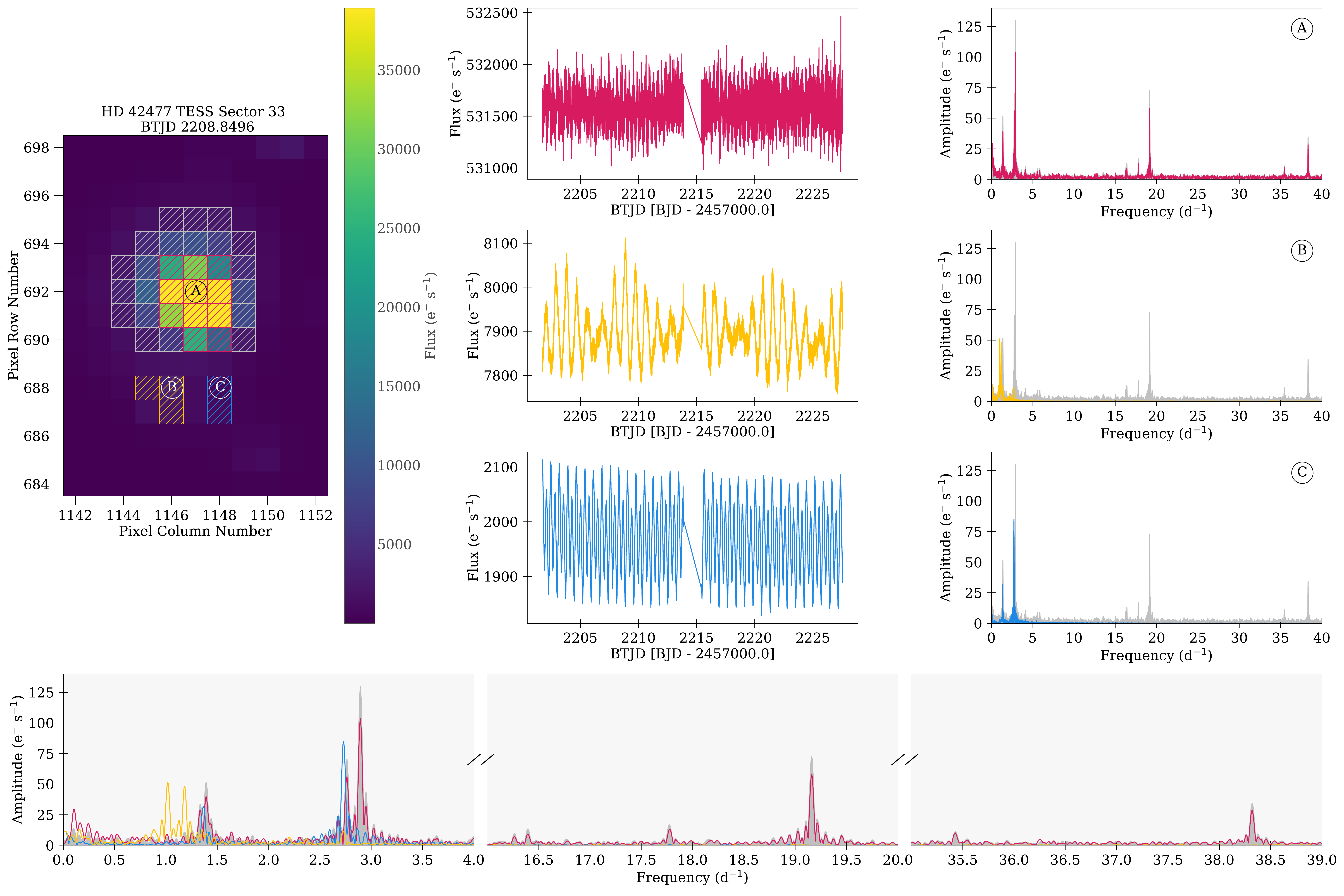} 
\caption{The results of the study of the variability content of the Target Pixel File of HD~42477. Left: the TPF with the aperture masks overplotted: White corresponds to the PDCSAP pipeline mask, while red corresponds to our custom mask for HD~42477. The other two masks indicated were used to generate custom light curves of potential contaminant stars -- orange for HD~253085, and blue for Gaia160. Middle and right: colour-coded light curves created using the corresponding masks corrected with Cotrending Basis Vectors and their amplitude spectra. HD~42477 is plotted in the top panel (A), HD~253085 in the middle panel (B) and Gaia160 in the bottom panel (C). The PDCSAP  amplitude spectrum for HD\,42477 is shown in grey in each panel for comparison. Bottom: 3-panel zoom into the ranges of frequencies detected in HD~42477. See the text in Section \ref{sec:px-by-px} for more detail.}
\label{fig:contam}
\end{center}
\end{figure*}

While this was reassuring, we sought to ensure that there was zero contribution from other sources in the TPF to the observed photometric variability. To do so, we extracted a raw light curve for HD\,42477 using a much smaller aperture mask than the one used by the SPOC pipeline (31 pixels in the pipeline mask versus 11 in our chosen mask) to minimise any possible contamination. These differing masks are shown in Fig.\,\ref{fig:contam} in red and white shading. It is important to note that HD~42477, with TESS magnitude $T = 6.01$, is beyond the saturation level of TESS, which is about $T \leq 6.8$ \citep{2015ApJ...809...77S}. The CCDs in the TESS camera are able to conserve charge from even very saturated stars by spreading the excess charge across adjacent pixels in the CCD column, which we took into account when creating a new aperture mask for the target. 

We identified two variable sources in the TPF at positions corresponding to stars number 5 and 7 in Fig.\,\ref{fig:tpfgaia}. Star 5 is a $G=10.34$~mag F2V star, with the designation HD~253085. Star number 7 is a $G=11.90$~mag star, with designation Gaia DR3 3344372393516924160 (hereafter referred to as Gaia160). Gaia160 has an unknown spectral type; however, \citet{2019AJ....158...93B} computed $T_{\rm eff} = 7700\pm 400$\,K. On the other hand, Gaia DR3 provides a $T_{\rm eff} = 9570^{+65}_{-60}$\,K, with a $BP-RP$ colour of 0.58 \citep{GAIA2016, gaia-dr3-paper}. The colour agrees with the $T_{\rm eff}$ presented in \citeauthor{2019AJ....158...93B}, suggesting a late-A or early-F star based on the tables of \citet{2013ApJS..208....9P}; the $T_{\rm eff}$ from Gaia does not agree with this interpretation, however, and instead suggests an early A star. Nevertheless, the key result is that it probably is in the instability strip and could potentially contaminate the light curve of HD\,42477 with putative high-amplitude pulsations. Finally, there is a $G=13$ star almost exactly between Gaia160 and HD\,253085 (star 9 in Fig.\,\ref{fig:tpfgaia}, with designation Gaia EDR3 3344373802266195200, which we abbreviate to Gaia200). This star was also investigated for any possible pulsations, but due to its comparatively faint magnitude ($\Delta G_{\rm mag}\sim 7$), we focused on the other two stars for our analysis.

To characterize the variability of these potential contaminants, we extracted raw light curves for these stars using the apertures shown in Fig.\,\ref{fig:contam} and corrected them using Cotrending Basis Vectors within \texttt{lightkurve}. The light curves for HD\,42477, HD\,253085 and Gaia160 are shown in the three middle panels of Fig.\,\ref{fig:contam}: HD~42477 in red, calculated with aperture (A); HD~253085 in orange, calculated with aperture (B); and Gaia160 in blue, calculated using aperture (C). The right-hand panels show the amplitude spectra of these light curves between 0 to 40~d$^{-1}$, with that calculated for HD\,42477 using the PDCSAP light curve (aperture shown in white) shown in grey.

From this analysis, it is evident that the high frequencies beyond 15~d$^{-1}$ clearly originate in HD~42477 only, while low-frequency variability is present in all three stars. The bottom three panels of Fig.\,\ref{fig:contam} show a zoom into the frequency ranges of interest. In the low frequency range, it is evident that the frequencies are very close. The two highest amplitude frequencies of HD~253085 are $1.018325(93)$~d$^{-1}$ and $1.180372(96)$~d$^{-1}$, and they do not overlap with any of the frequencies of HD~42477. For Gaia160 the two highest amplitude frequencies are $2.729944(21)$~d$^{-1}$ and $1.365009(56)$~d$^{-1}$, which appear to be harmonically related. These two are close to $\nu_{2}$ and $\nu_{1}$ of HD~42477, respectively, but still are independent frequencies, given that they are separated by more than the Rayleigh frequency resolution of the entire data set. Finally, while this is not shown in Fig.\,\ref{fig:contam}, Gaia200 exhibits a significant peak at 4.92614(43) d$^{-1}$, which is not near any of the other frequencies we have investigated. This suggests it is unlikely to affect our analysis. From this, we can conclusively state that the p~mode ($\nu_4$) originates from HD\,42477 and that the nearby stars (which show their own intrinsic variability) are not the source of any of the observed low frequencies ($\nu_1$, $\nu_2$, $\nu_3$) that make up the combination frequencies shown in Table \ref{Tab:nld}.

\section{Spectroscopic search for a binary companion}
\label{sec:specsearch}

We searched for the presence of spectral lines of a binary companion to allow us to test whether the pulsations observed in the TESS light curve of HD~42477 originate in that star or in a fainter A-F star binary companion. Such a star could itself show p~, g~ and r~modes. Because Be stars are thought to have been spun up by mass transfer from a companion at an earlier stage of their evolution, we do not expect a close main-sequence companion. However, a wide main-sequence companion that would not have influenced or been perturbed by earlier mass transfer in an inner pair needs to be ruled out. The presence of any compact companion, such as a white dwarf, is not relevant in this context, as such a star would be too faint to contribute to the total light and would not have the observed pulsation frequencies. We additionally note that the Gaia DR3 renormalised unit weight error (RUWE) value for HD~42477 is 1.079, so this star is not an obvious astrometric binary.

Spectral regions including prominent absorption lines (e.g., \ion{Fe}{ii} 5169\,\AA\,and H$\beta$) in the NRES spectra were searched for the presence of line profile variations by phasing the mean-subtracted residuals to $\nu_1$, $\nu_2$ and $\nu_3$. However, with only relatively few observations, this test was inconclusive. The spectra were also compared to synthetic spectra to identify the signature of a potential companion; these were computed using the {\tt SPECTRUM} code \citep{1994AJ....107..742G}, with the parameters $T_{\rm eff}=10\,000$~K, log~$g=3.5$ and $v\,\sin\,i = 220$\,km~s$^{-1}$. The effective temperature was chosen to be consistent with that used for modelling in Section\,\ref{sec:model} below. We verified the choice of $\log\,g$ by comparing the observed wings of the hydrogen line profiles with synthetic ones and found the best fit to occur at $\log\,g=3.5$; similarly, our choice of $v\,\sin\,i$ was based on a comparison to observed spectra.

We then added synthetic spectra of three hypothetical companions to the spectrum of the primary. Companion 1 was assumed to resemble a slowly-rotating main-sequence star of spectral type F0 ($T_{\rm eff}=7250$~K, log~$g=4.25$, $v\,\sin\,i = 30$\,km~s$^{-1}$). Companion 2 had the same $T_{\rm eff}$ and log~$g$, but rotated more rapidly ($v \sin i = 100$\,km~s$^{-1}$), whereas Companion 3 was chosen to match an even more rapidly-rotating F3V star ($T_{\rm eff}=6750$~K, log~$g=4.25$, $v\,\sin\,i = 200$\,km~s$^{-1}$). After computing these synthetic spectra, we tested the magnitude difference required to be able to detect such companions in the observed spectra.

We found that Companion 1 would be detectable even if it was 3.5 magnitudes fainter in the $V$ band than the primary star, owing to its sharp spectral lines that are in stark contrast to those of the primary. Companion 2 would be detectable at a $V$ magnitude difference of no more than 2.5, while Companion 3, however, would escape detection by only being about 1.25 magnitudes fainter in $V$. This suggests that one of the main factors influencing a putative companion's detectability is its projected rotation rate.

Using the absolute magnitude of HD~42477 as determined in Section\,\ref{sec:model}, in conjunction with the tables of \cite{2013ApJS..208....9P}, we estimate that a physical F0V companion would be some $1.8 -2.1$ mag fainter in $V$. The ultimate conclusion from this part of our study, then, is that our spectra are sufficient to exclude a physical companion to HD~42477 of early to mid-F spectral type if it rotates slower than $v\,\sin\,i\sim100$\,km~s$^{-1}$ , but we would not be able to detect a physically associated rapidly-rotating companion or an optical companion. We can largely rule out a star blended into the TESS pixels of our target at a larger distance based on Gaia data and our analysis in Section \ref{sec:px-by-px}.

\section{Modelling HD~42477}
\label{sec:model}

\subsection{Global parameters}

To obtain initial values of $T_{\rm eff}$, $\log g$, luminosity and radius to model HD~42477, we used a variety of existing photometric observations, including Str\"omgren $uvby$ plus H$\beta$  and Johnson $BV$ photometry.

The Str\"omgren indices are $b-y = 0.026$, $m_1 = 0.128$, $c_1 = 1.126$ and ${\rm H}\beta = 2.820$. Using the B star calibration provided in Figure 3 of \citet{1985MNRAS.217..305M} yields $T_{\rm eff} = 9900$\,K and $\log g = 3.4$. Using the A star calibration from Figure 2 of the same work yields $T_{\rm eff} = 7900$\,K and $\log g = 3.3$. The reason for this ambiguity is that the H$\beta$ index peaks at A0, where the Balmer lines have maximum strength (by the definition of the spectral types). So, H$\beta$ drops from its peak on either side of A0. We can use the value of the index $b-y$ to decide that the hotter $T_{\rm eff}$ is correct; this value is calibrated to be 0.0 at A0. Using the calibration presented in Tables 1 and 2 of \citet{1979AJ.....84.1858C} gives $b-y = 0.126$ for H$\beta = 2.82$ and a spectral type of A7 for H$\beta = 2.824$, respectively. His table suggests a spectral type of A1 or A2 for $b-y = 0.026$.\footnote{Note that these indices include reddening due to interstellar dust.} As a result, we confirm the early A classification and also find that the star is located above the Zero Age Main Sequence (ZAMS), where $\log g = 4.5$ for A0 stars.   

From the Gaia DR3 \citep{GAIA2016,gaia-dr3-paper} parallax, $6.427 \pm 0.050$\,mas, we obtain  a distance of $155.6 \pm 1.2$\,pc,  yielding a distance modulus $\mu = 5.960\pm 0.017$. Photometric measurements of this star using Johnson filters in \citet{2000A&A...355L..27H} give $V=  6.036$, $B= 6.050$ and $B-V= 0.014$. Using the dust maps of \citet{Lallement2018}, we find a colour excess of $E(B-V) = 0.007$ for the direction and at the distance of HD~42477. This results in $(B-V)_0 = 0.007$, consistent with the A0Vnne spectral classification provided in \citet{1969AJ.....74..375C}.
Using the relation between $B-V$ and $T_{\rm eff}$ and the Bolometric Correction (BC) as given in \citet{1996ApJ...469..355F}, we find $\log T_{\rm eff}= 3.9747,  T_{\rm eff} = 9434$\,K,  and ${\rm BC} = -0.134$. 

Thus, the absolute bolometric magnitude is $M_{\rm bol} = -0.080$,  corresponding to $\log L/{\rm L}_\odot = 1.932$. We rearrange the Stefan-Boltzmann relation, $L = 4 \pi  R^2 \sigma T^4$, and normalise it to the Sun. This yields 
\begin{equation}
    \log L/{\rm L}_\odot = 2\log R/{\rm R}_\odot + 4(\log T_{\rm eff}-3.76185),
\end{equation}
from which we obtain $\log R/{\rm R}_\odot = 0.540$, or $R =  3.5$\,R$_\odot$. These global parameters are consistent with a somewhat evolved $2.8$-M$_\odot$ main-sequence model of $(X,Z)=(0.72,0.014).$ 
We have selected a model (Table~\ref{tab:models}) from the main-sequence evolutionary track of $2.8$-M$_\odot$ stars that has parameters consistent with HD~42477.
Main-sequence evolution models (without rotation) were obtained using the MESA code \citep[v.7184;][]{pax11,pax13,pax15}, in which the convective core boundary was determined by the Schwarzschild criterion, elemental diffusion was activated to have a smooth Brunt-V\"ais\"al\"a frequency, and radiation turbulence was also activated to prevent helium settling.

\begin{table}
\centering
\caption{{Model parameters for the best-fit model for HD\,42477}}
\begin{tabular}{ccccc}
\hline
$M/{\rm M}_\odot$ & $\log L/{\rm L}_\odot$ & $\log T_{\rm eff}$ & $\log R/{\rm R}_\odot$ & $\log g$ \\
\hline
2.8 & $1.940$ & $3.9928$ & $0.508$ & $3.87$\\
\hline
\end{tabular}
\label{tab:models}
\end{table}

\subsection{r modes}

r~mode oscillations are global Rossby waves influenced by buoyancy. 
These oscillations are predominantly horizontal, and the pattern propagates in the direction opposite to the rotation in the co-rotating frame; they are seen propagating in the same direction of rotation in the inertial frame, however, because the phase speed in the co-rotating frame is slower than the rotation. For this reason, r~modes with azimuthal order $m \ge 1$ have frequencies in the inertial frame between $(m-1)\nu_{\rm rot}$ and $m\nu_{\rm rot}$    
\citep[see e.g.,][for details]{2018MNRAS.474.2774S}.\footnote{In this paper we adopt the convention that $m>0$ and $m<0$ indicate, respectively, retrograde and prograde (in the co-rotating frame) modes.}

\begin{figure}
  \includegraphics[width=0.5\textwidth]{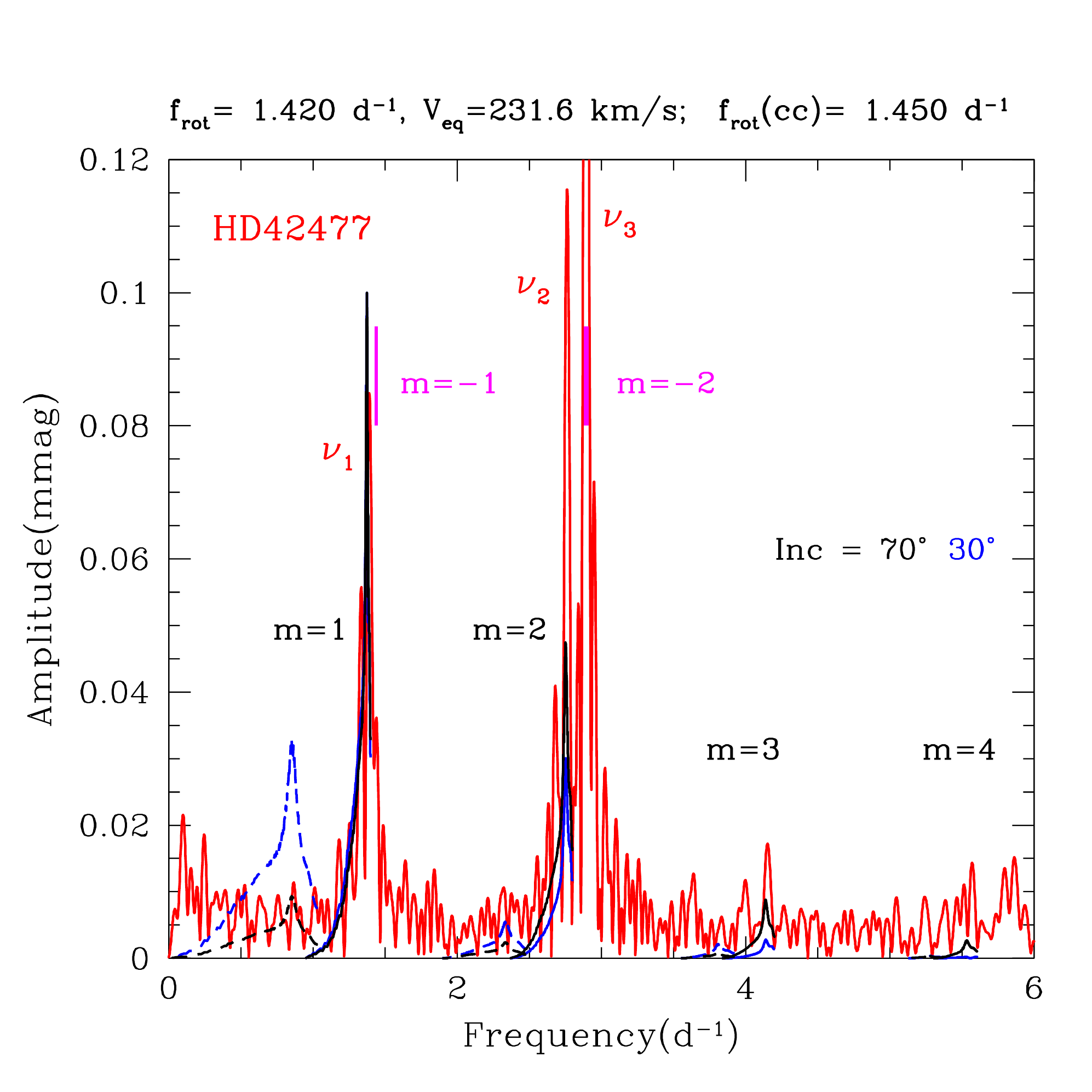} 
\caption{The visibility distributions of r modes (blue and black lines for assumed inclinations of $30^\circ$ and $70^\circ$, respectively) are compared with the observed Sector 33 amplitude spectrum of HD\,42477 (red line). The visibility is obtained under the assumption of energy equipartition, and an arbitrary normalisation of the maximum of the $m=1$ modes to be $0.1$~mmag. Solid and dashed lines are for r~modes whose temperature variations on the surface are, respectively, symmetric and anti-symmetric with respect to the equator. Short vertical magenta lines indicate frequencies of $m=-1$ and $-2$ prograde g~modes excited by an overstable convective (OsC) mode.}
\label{fig:rmodes}
\end{figure}

If we adopt a rotation frequency of $1.42$~d$^{-1}$, the low frequency features of  HD~42477 can be fitted with r-mode frequency ranges reasonably well. Fig.\,\ref{fig:rmodes} shows the expected visibility distributions of r~modes, which were computed using the Traditional Approximation of Rotation as discussed in \citet{2018MNRAS.474.2774S}, when the star is seen with inclinations $i$ (i.e., the angle between the rotational axis and the line of sight) of $70^\circ$ (black lines) and $30^\circ$ (blue lines; solid and dashed lines are for symmetric and anti-symmetric modes with respect to the equator).\footnote{The actual inclination is probably between these based on the morphology of the weak H$\alpha$ emission,  i.e., the inclination is neither very high nor very low.} The kinetic energy of the r~modes is assumed to be distributed equally, and the visibility is normalised arbitrarily to 0.1~mmag at the peak of the $m=1$ r~modes. 

If energy is equally distributed to r~modes, the visibilities of symmetric $m=1$ modes should be largest. However, the peaks around $2.5 - 2.8$\,d$^{-1}$ of HD~42477 indicate that a much larger amount of energy  ($\sim$$3^2$ times as much) is given to $m=2$ modes. These modes show amplitude modulation on the timescale of months, as discussed in Section \ref{sec:fa1} and shown in Fig.\,\ref{fig:ft2}. This analysis was done using only the Sector 33 data with fixed amplitudes, so it does not account for the changes in mode amplitude observed in Sectors 43-45.

\subsection{Overstable convection modes}

\citet{2020MNRAS.497.4117L} found that overstable convective core modes excite high order g~mode(s) in the envelope, which should be observed as rotational modulations associated with the rotation frequency of the core, i.e., $\sim |m|\nu_{\rm rot}$(cc), where (cc) refers to the convective core.
If the rotation frequency of the convective core is assumed to be $1.45~$d$^{-1}$  (and the envelope rotates at $1.42~$d$^{-1}$),  an overstable convective mode of $m=-2$ excites envelope g~modes at $\sim2.89$\,d$^{-1}$, {while an $m=-1$ convective mode excites g~modes at $\sim1.44$\,d$^{-1}$}.
These frequencies are shown by short vertical magenta lines in Fig.\,\ref{fig:rmodes} (frequency differences among these g~modes with a given azimuthal order $m$ are too small to be resolved). The $m=-2$ modes are consistent with $\nu_3=2.89362$\,d$^{-1}$ of HD\,42477, while no obvious peaks are present corresponding to the $m=-1$ modes. The reason for the difference may be attributed to the fact that typical growth rates of the latter are about ten times smaller than the former.

Using a local mixing length theory, we find the convective turnover time to be between $30 - 40$\,d in most parts of the convective core. This suggests that convective motions in the core may be responsible for the amplitude modulations of low-frequency peaks of HD 42477 across TESS sectors, as seen in Fig.\,\ref{fig:ft2}.

Since overstable convective (OsC) modes in resonance with g~modes have amplitudes both in the core and in the envelope, they play a role in angular momentum transfer between the two regions. Prograde OsC modes excited in the core extract angular momentum from the rotating fluid and deposit excess angular momentum to fluids in the envelope, where the modes are damped by radiative dissipation. \citet{2021MNRAS.505.1495L} computed OsC modes of 2-M$_\odot$, 4-M$_\odot$ and 20-M$_\odot$ main-sequence stars and suggested that the OsC modes can transport angular momentum from the convective core to the envelope by resonantly exciting envelope g~modes. \citeauthor{2021MNRAS.505.1495L} also found that for OsC modes, the timescale of angular momentum transfer $\tau$ in the core is much longer than that in the surface layers. For example, with the results for his rapidly rotating 2-M$_\odot$ ZAMS model, assuming a luminosity amplitude $|\delta L_{\rm R}/L_{\rm R}|\sim 0.3\times10^{-3}$ (as suggested by Fig.\,\ref{fig:ft1}), we obtain $\tau\sim 10^9$ yr in the core and $\tau\sim 10^6$ yr at the surface. This means that the timescale $\tau$ at the surface is significantly shorter than the lifetime of a 2-M$_\odot$ main sequence star. 

In general, as the mass of the star increases, $\tau$ becomes shorter. For a 20-M$_\odot$ ZAMS star, assuming the same value as above for $|\delta L_{\rm R}/L_{\rm R}|$, we obtain $\tau\sim 10^8$ yr in the core and $\tau\sim 5\times 10^4$ yr at the surface. For the oscillation amplitudes at the surface that we assume, $\tau$ in the core is shorter than the lifetime of a 20-M$_\odot$ main sequence star. Note that $\tau$ is inversely proportional to $(\delta L_{\rm R}/L_{\rm R})^2$, the square of the luminosity amplitude.

\subsection{p modes}\label{sec:pmodes}

Since HD~42477 is hotter than the blue boundary of the $\delta$~Sct instability strip, no p modes in this star are thought to be excited by the $\kappa$ mechanism in the \ion{He}{ii} ionization zone. This disagrees with our findings from the TESS photometry for this star, which shows a p~mode. The stability of p modes was calculated using a nonadiabatic pulsation code based on the method of \citet{1995A&A...301..419L}, in which we include rotational effects on pulsation by expanding the eigenfunction as a sum of terms proportional to spherical harmonics. Our spherical models do not show excitation of any p~modes, as expected. This discrepancy could be resolved if the equatorial region is  deformed (and cooled) due to the centrifugal force.

However, the detected p~mode is clearly coupled to the g~modes (as shown by the combination frequencies); we suggest that these g~modes are themselves excited by OsC core modes. We conjecture that the single observed p~mode is driven by coupling with the g~modes, or that the oblateness of this rapidly rotating star permits driving by \ion{He}{ii} ionization in the equatorial region. Both of these are possible explanations for the observed p~mode excitation in stars between the $\beta$\,Cep and $\delta$~Sct instability strips. 

\section{conclusions}

We set out to disprove the conjecture that the p-mode pulsators that appear to lie between the $\beta$\,Cep and $\delta$~Sct instability strips are all the result of incorrect determinations of $T_{\rm eff}$, or contamination by a background star or orbital companion. We performed a thorough study of potentially contaminating background stars and also searched for an orbital companion in the spectrum of HD~42477, an A0Vnne star for which the TESS photometry shows a single definite p~mode coupled to lower frequency g~modes. No previous study has ruled out the possibility of an orbital companion as the source of the p~mode(s) in any star in the range between these two p-mode instability strips. We have essentially succeeded, leaving only the unlikely possibility of a rapidly rotating, early F-type companion as the source of the pulsations. The typical $\delta$~Sct star with these characteristics demonstrates multiple p~modes; a single such mode in a star of this kind is improbable. And, as we pointed out in Section\,\ref{sec:background}, virtually all Be stars also show g~modes, making it even more improbable that the g~modes could be in a companion to HD~42477.

Given that a rapidly rotating, early F, singly-periodic $\delta$~Sct companion to HD~42477 is highly unlikely, and given that it is unlikely that an A0Vnne star would not show any g~modes, 
we have demonstrated that one star in the temperature range between the $\beta$\,Cep and $\delta$~Sct stars pulsates with a p~mode. This disproves the conjecture that all such stars might be explained by temperature uncertainties, binary companions, or contamination by background pulsators. This is the first time that such proof has been given. Given the fraction of such stars found in this range by  \citet{2009A&A...506..471D}, \citet{2018MNRAS.476.3169B} and by the examination of thousands of B and A stars by \citet{2020MNRAS.493.5871B}, we concur that p~mode pulsators can exist in the $T_{\rm eff}$ range between the $\beta$\,Cep and $\gamma$~Dor instability strips. Moreover, our modelling shows the presence of both g~modes and r~modes in HD~42477, thus demonstrating g~mode pulsation in a star cooler than the SPB theoretical instability strip. While it has not been shown yet, visual inspection of {\it Kepler} and TESS light curves of Be stars indicates that r~modes are often excited in these stars. 

The coupling of the g~modes and p~mode in HD~42477 and the theoretical understanding of overstable convective core modes as a source of excitation of g~modes (such as in the recent studies of \citealt{2020MNRAS.497.4117L} and \citealt{2021MNRAS.505.1495L}) suggest that mode coupling is important in understanding mode excitation. New calculations of the boundaries of the instability strips of $\beta$\,Cep, SPB, $\delta$~Sct and $\gamma$~Dor stars with the inclusion of mode coupling may elucidate the presence of pulsators between the currently defined instability strips. 
 
\section*{Acknowledgements}
The authors would like to thank Keaton Bell for illuminating discussions about the contamination analysis.

This work has been partially supported by the Polish National Science Center (NCN) grants 2015/18/A/ST9/00578 and  2021/43/B/ST9/02972 to GH. J.L.-B. acknowledges support from FAPESP (grant 2017/23731-1). This paper includes data collected by the TESS mission, specifically through Guest Investigator Programs G03186 (PI: Labadie-Bartz) and G04067 (PI: Wisniewski). Funding for TESS is provided by NASA's Science Mission Directorate. Resources used in this work were provided by the NASA High End Computing (HEC) Program through the NASA Advanced Supercomputing (NAS) Division at Ames Research Center for the production of the SPOC data products. 

This work has made use of observations from the LCOGT network, and also has made use of data from the European Space Agency (ESA) mission {\it Gaia} (\url{https://www.cosmos.esa.int/gaia}), processed by the {\it Gaia} Data Processing and Analysis Consortium (DPAC, \url{https://www.cosmos.esa.int/web/gaia/dpac/consortium}). Funding for the DPAC has been provided by national institutions, in particular the institutions participating in the {\it Gaia} Multilateral Agreement. This research made use of Lightkurve, a Python package for Kepler and TESS data analysis (Lightkurve Collaboration, 2018), and \texttt{tpfplotter} by J. Lillo-Box (publicly available in www.github.com/jlillo/tpfplotter), which also made use of the python packages \texttt{astropy}, \texttt{lightkurve}, \texttt{matplotlib} and \texttt{numpy}.

\section*{Data Availability}
The data underlying this article will be shared on reasonable request to the authors.

\bibliographystyle{mnras}
\bibliography{294125876}{}

\begin{thebibliography}{}
\makeatletter
\relax
\def\mn@urlcharsother{\let\do\@makeother \do\$\do\&\do\#\do\^\do\_\do\%\do\~}
\def\mn@doi{\begingroup\mn@urlcharsother \@ifnextchar [ {\mn@doi@}
  {\mn@doi@[]}}
\def\mn@doi@[#1]#2{\def\@tempa{#1}\ifx\@tempa\@empty \href
  {http://dx.doi.org/#2} {doi:#2}\else \href {http://dx.doi.org/#2} {#1}\fi
  \endgroup}
\def\mn@eprint#1#2{\mn@eprint@#1:#2::\@nil}
\def\mn@eprint@arXiv#1{\href {http://arxiv.org/abs/#1} {{\tt arXiv:#1}}}
\def\mn@eprint@dblp#1{\href {http://dblp.uni-trier.de/rec/bibtex/#1.xml}
  {dblp:#1}}
\def\mn@eprint@#1:#2:#3:#4\@nil{\def\@tempa {#1}\def\@tempb {#2}\def\@tempc
  {#3}\ifx \@tempc \@empty \let \@tempc \@tempb \let \@tempb \@tempa \fi \ifx
  \@tempb \@empty \def\@tempb {arXiv}\fi \@ifundefined
  {mn@eprint@\@tempb}{\@tempb:\@tempc}{\expandafter \expandafter \csname
  mn@eprint@\@tempb\endcsname \expandafter{\@tempc}}}

\bibitem[\protect\citeauthoryear{{Aerts}}{{Aerts}}{2021}]{2021RvMP...93a5001A}
{Aerts} C.,  2021, \mn@doi [Reviews of Modern Physics]
  {10.1103/RevModPhys.93.015001}, \href
  {https://ui.adsabs.harvard.edu/abs/2021RvMP...93a5001A} {93, 015001}

\bibitem[\protect\citeauthoryear{{Aerts}, {Christensen-Dalsgaard}  \&
  {Kurtz}}{{Aerts} et~al.}{2010}]{2010aste.book.....A}
{Aerts} C.,  {Christensen-Dalsgaard} J.,   {Kurtz} D.~W.,  2010,
  {Asteroseismology}.
Astronomy and Astrophysics Library. Springer Science+Business Media B.V.

\bibitem[\protect\citeauthoryear{{Aller}, {Lillo-Box}, {Jones}, {Miranda}  \&
  {Barcel{\'o} Forteza}}{{Aller} et~al.}{2020}]{2020A&A...635A.128A}
{Aller} A.,  {Lillo-Box} J.,  {Jones} D.,  {Miranda} L.~F.,   {Barcel{\'o}
  Forteza} S.,  2020, \mn@doi [\aap] {10.1051/0004-6361/201937118}, \href
  {https://ui.adsabs.harvard.edu/abs/2020A&A...635A.128A} {635, A128}

\bibitem[\protect\citeauthoryear{{Baade} et~al.,}{{Baade}
  et~al.}{2018a}]{2018pas8.conf...69B}
{Baade} D.,  et~al., 2018a, in {Wade} G.~A.,  {Baade} D.,  {Guzik} J.~A.,
  {Smolec} R.,  eds,  Vol. 8, 3rd BRITE Science Conference. pp 69--76
  (\mn@eprint {arXiv} {1708.08413})

\bibitem[\protect\citeauthoryear{{Baade} et~al.,}{{Baade}
  et~al.}{2018b}]{2018A&A...610A..70B}
{Baade} D.,  et~al., 2018b, \mn@doi [\aap] {10.1051/0004-6361/201731187}, \href
  {http://esoads.eso.org/abs/2018A%26A...610A..70B} {610, A70}

\bibitem[\protect\citeauthoryear{{Bai}, {Liu}, {Bai}, {Wang}  \& {Fan}}{{Bai}
  et~al.}{2019}]{2019AJ....158...93B}
{Bai} Y.,  {Liu} J.,  {Bai} Z.,  {Wang} S.,   {Fan} D.,  2019, \mn@doi [\aj]
  {10.3847/1538-3881/ab3048}, \href
  {https://ui.adsabs.harvard.edu/abs/2019AJ....158...93B} {158, 93}

\bibitem[\protect\citeauthoryear{{Balona} \& {Ozuyar}}{{Balona} \&
  {Ozuyar}}{2020}]{2020MNRAS.493.5871B}
{Balona} L.~A.,  {Ozuyar} D.,  2020, \mn@doi [\mnras] {10.1093/mnras/staa670},
  \href {https://ui.adsabs.harvard.edu/abs/2020MNRAS.493.5871B} {493, 5871}

\bibitem[\protect\citeauthoryear{{Bowman} \& {Kurtz}}{{Bowman} \&
  {Kurtz}}{2018}]{2018MNRAS.476.3169B}
{Bowman} D.~M.,  {Kurtz} D.~W.,  2018, \mn@doi [\mnras] {10.1093/mnras/sty449},
  \href {https://ui.adsabs.harvard.edu/abs/2018MNRAS.476.3169B} {476, 3169}

\bibitem[\protect\citeauthoryear{{Bowman} et~al.,}{{Bowman}
  et~al.}{2019}]{2019NatAs...3..760B}
{Bowman} D.~M.,  et~al., 2019, \mn@doi [Nature Astronomy]
  {10.1038/s41550-019-0768-1}, \href
  {https://ui.adsabs.harvard.edu/abs/2019NatAs...3..760B} {3, 760}

\bibitem[\protect\citeauthoryear{{Bowman}, {Burssens}, {Sim{\'o}n-D{\'\i}az},
  {Edelmann}, {Rogers}, {Horst}, {R{\"o}pke}  \& {Aerts}}{{Bowman}
  et~al.}{2020}]{2020A&A...640A..36B}
{Bowman} D.~M.,  {Burssens} S.,  {Sim{\'o}n-D{\'\i}az} S.,  {Edelmann}
  P.~V.~F.,  {Rogers} T.~M.,  {Horst} L.,  {R{\"o}pke} F.~K.,   {Aerts} C.,
  2020, \mn@doi [\aap] {10.1051/0004-6361/202038224}, \href
  {https://ui.adsabs.harvard.edu/abs/2020A&A...640A..36B} {640, A36}

\bibitem[\protect\citeauthoryear{{Brown} et~al.,}{{Brown}
  et~al.}{2013}]{2013PASP..125.1031B}
{Brown} T.~M.,  et~al., 2013, \mn@doi [\pasp] {10.1086/673168}, \href
  {https://ui.adsabs.harvard.edu/abs/2013PASP..125.1031B} {125, 1031}

\bibitem[\protect\citeauthoryear{{Cowley}, {Cowley}, {Jaschek}  \&
  {Jaschek}}{{Cowley} et~al.}{1969}]{1969AJ.....74..375C}
{Cowley} A.,  {Cowley} C.,  {Jaschek} M.,   {Jaschek} C.,  1969, \mn@doi [\aj]
  {10.1086/110819}, \href
  {https://ui.adsabs.harvard.edu/abs/1969AJ.....74..375C} {74, 375}

\bibitem[\protect\citeauthoryear{{Crawford}}{{Crawford}}{1979}]{1979AJ.....84.1858C}
{Crawford} D.~L.,  1979, \mn@doi [\aj] {10.1086/112617}, \href
  {https://ui.adsabs.harvard.edu/abs/1979AJ.....84.1858C} {84, 1858}

\bibitem[\protect\citeauthoryear{{Degroote} et~al.,}{{Degroote}
  et~al.}{2009}]{2009A&A...506..471D}
{Degroote} P.,  et~al., 2009, \mn@doi [\aap] {10.1051/0004-6361/200911884},
  \href {https://ui.adsabs.harvard.edu/abs/2009A&A...506..471D} {506, 471}

\bibitem[\protect\citeauthoryear{{Dupret}, {Grigahc{\`e}ne}, {Garrido},
  {Gabriel}  \& {Scuflaire}}{{Dupret} et~al.}{2004}]{2004A&A...414L..17D}
{Dupret} M.~A.,  {Grigahc{\`e}ne} A.,  {Garrido} R.,  {Gabriel} M.,
  {Scuflaire} R.,  2004, \mn@doi [\aap] {10.1051/0004-6361:20031740}, \href
  {https://ui.adsabs.harvard.edu/abs/2004A&A...414L..17D} {414, L17}

\bibitem[\protect\citeauthoryear{{Dupret}, {Grigahc{\`e}ne}, {Garrido},
  {Gabriel}  \& {Scuflaire}}{{Dupret} et~al.}{2005}]{2005A&A...435..927D}
{Dupret} M.~A.,  {Grigahc{\`e}ne} A.,  {Garrido} R.,  {Gabriel} M.,
  {Scuflaire} R.,  2005, \mn@doi [\aap] {10.1051/0004-6361:20041817}, \href
  {https://ui.adsabs.harvard.edu/abs/2005A&A...435..927D} {435, 927}

\bibitem[\protect\citeauthoryear{{Flower}}{{Flower}}{1996}]{1996ApJ...469..355F}
{Flower} P.~J.,  1996, \mn@doi [\apj] {10.1086/177785}, \href
  {https://ui.adsabs.harvard.edu/abs/1996ApJ...469..355F} {469, 355}

\bibitem[\protect\citeauthoryear{{Gaia Collaboration} et~al.,}{{Gaia
  Collaboration} et~al.}{2016}]{GAIA2016}
{Gaia Collaboration} et~al., 2016, \mn@doi [\aap]
  {10.1051/0004-6361/201629272}, \href
  {https://ui.adsabs.harvard.edu/abs/2016A&A...595A...1G} {595, A1}

\bibitem[\protect\citeauthoryear{{Gaia Collaboration} et~al.,}{{Gaia
  Collaboration} et~al.}{2022}]{gaia-dr3-paper}
{Gaia Collaboration} et~al., 2022, arXiv e-prints, \href
  {https://ui.adsabs.harvard.edu/abs/2022arXiv220800211G} {p. arXiv:2208.00211}

\bibitem[\protect\citeauthoryear{{Gray} \& {Corbally}}{{Gray} \&
  {Corbally}}{1994}]{1994AJ....107..742G}
{Gray} R.~O.,  {Corbally} C.~J.,  1994, \mn@doi [\aj] {10.1086/116893}, \href
  {https://ui.adsabs.harvard.edu/abs/1994AJ....107..742G} {107, 742}

\bibitem[\protect\citeauthoryear{{Grigahc{\`e}ne}, {Dupret}, {Gabriel},
  {Garrido}  \& {Scuflaire}}{{Grigahc{\`e}ne}
  et~al.}{2005}]{2005A&A...434.1055G}
{Grigahc{\`e}ne} A.,  {Dupret} M.~A.,  {Gabriel} M.,  {Garrido} R.,
  {Scuflaire} R.,  2005, \mn@doi [\aap] {10.1051/0004-6361:20041816}, \href
  {https://ui.adsabs.harvard.edu/abs/2005A&A...434.1055G} {434, 1055}

\bibitem[\protect\citeauthoryear{{Grigahc{\`e}ne} et~al.,}{{Grigahc{\`e}ne}
  et~al.}{2010}]{2010ApJ...713L.192G}
{Grigahc{\`e}ne} A.,  et~al., 2010, \mn@doi [\apjl]
  {10.1088/2041-8205/713/2/L192}, \href
  {https://ui.adsabs.harvard.edu/abs/2010ApJ...713L.192G} {713, L192}

\bibitem[\protect\citeauthoryear{{H{\o}g} et~al.,}{{H{\o}g}
  et~al.}{2000}]{2000A&A...355L..27H}
{H{\o}g} E.,  et~al., 2000, \aap, \href
  {https://ui.adsabs.harvard.edu/abs/2000A&A...355L..27H} {355, L27}

\bibitem[\protect\citeauthoryear{{Houdek} \& {Dupret}}{{Houdek} \&
  {Dupret}}{2015}]{2015LRSP...12....8H}
{Houdek} G.,  {Dupret} M.-A.,  2015, \mn@doi [Living Reviews in Solar Physics]
  {10.1007/lrsp-2015-8}, \href
  {https://ui.adsabs.harvard.edu/abs/2015LRSP...12....8H} {12, 8}

\bibitem[\protect\citeauthoryear{{Jenkins} et~al.,}{{Jenkins}
  et~al.}{2016}]{jenkins-spoc}
{Jenkins} J.~M.,  et~al., 2016, in {Chiozzi} G.,  {Guzman} J.~C.,  eds,
  Society of Photo-Optical Instrumentation Engineers (SPIE) Conference Series
  Vol. 9913, Software and Cyberinfrastructure for Astronomy IV. p. 99133E,
  \mn@doi{10.1117/12.2233418}

\bibitem[\protect\citeauthoryear{{Kahraman Ali{\c{c}}avu{\c{s}}}, {Poretti},
  {Catanzaro}, {Smalley}, {Niemczura}, {Rainer}  \& {Handler}}{{Kahraman
  Ali{\c{c}}avu{\c{s}}} et~al.}{2020}]{2020MNRAS.493.4518K}
{Kahraman Ali{\c{c}}avu{\c{s}}} F.,  {Poretti} E.,  {Catanzaro} G.,  {Smalley}
  B.,  {Niemczura} E.,  {Rainer} M.,   {Handler} G.,  2020, \mn@doi [\mnras]
  {10.1093/mnras/staa399}, \href
  {https://ui.adsabs.harvard.edu/abs/2020MNRAS.493.4518K} {493, 4518}

\bibitem[\protect\citeauthoryear{{Kochukhov}, {Piskunov}, {Sachkov}  \&
  {Kudryavtsev}}{{Kochukhov} et~al.}{2005}]{2005A&A...439.1093K}
{Kochukhov} O.,  {Piskunov} N.,  {Sachkov} M.,   {Kudryavtsev} D.,  2005,
  \mn@doi [\aap] {10.1051/0004-6361:20053123}, \href
  {https://ui.adsabs.harvard.edu/abs/2005A&A...439.1093K} {439, 1093}

\bibitem[\protect\citeauthoryear{{Kurtz}}{{Kurtz}}{1985}]{1985MNRAS.213..773K}
{Kurtz} D.~W.,  1985, \mn@doi [\mnras] {10.1093/mnras/213.4.773}, \href
  {https://ui.adsabs.harvard.edu/abs/1985MNRAS.213..773K} {213, 773}

\bibitem[\protect\citeauthoryear{{Kurtz}}{{Kurtz}}{2022}]{2022ARA&A..60...31K}
{Kurtz} D.~W.,  2022, \mn@doi [\araa] {10.1146/annurev-astro-052920-094232},
  \href {https://ui.adsabs.harvard.edu/abs/2022ARA&A..60...31K} {60, 31}

\bibitem[\protect\citeauthoryear{{Kurtz}, {Shibahashi}, {Murphy}, {Bedding}  \&
  {Bowman}}{{Kurtz} et~al.}{2015}]{2015MNRAS.450.3015K}
{Kurtz} D.~W.,  {Shibahashi} H.,  {Murphy} S.~J.,  {Bedding} T.~R.,   {Bowman}
  D.~M.,  2015, \mn@doi [\mnras] {10.1093/mnras/stv868}, \href
  {https://ui.adsabs.harvard.edu/abs/2015MNRAS.450.3015K} {450, 3015}

\bibitem[\protect\citeauthoryear{{Labadie-Bartz}, {Carciofi}, {Henrique de
  Amorim}, {Rubio}, {Luiz Figueiredo}, {Ticiani dos Santos}  \&
  {Thomson-Paressant}}{{Labadie-Bartz} et~al.}{2022}]{Labadie-Bartz2022}
{Labadie-Bartz} J.,  {Carciofi} A.~C.,  {Henrique de Amorim} T.,  {Rubio} A.,
  {Luiz Figueiredo} A.,  {Ticiani dos Santos} P.,   {Thomson-Paressant} K.,
  2022, \mn@doi [\aj] {10.3847/1538-3881/ac5abd}, \href
  {https://ui.adsabs.harvard.edu/abs/2022AJ....163..226L} {163, 226}

\bibitem[\protect\citeauthoryear{{Lallement} et~al.,}{{Lallement}
  et~al.}{2018}]{Lallement2018}
{Lallement} R.,  et~al., 2018, \mn@doi [\aap] {10.1051/0004-6361/201832832},
  \href {https://ui.adsabs.harvard.edu/abs/2018A&A...616A.132L} {616, A132}

\bibitem[\protect\citeauthoryear{{Lee}}{{Lee}}{2021}]{2021MNRAS.505.1495L}
{Lee} U.,  2021, \mn@doi [\mnras] {10.1093/mnras/stab1433}, \href
  {https://ui.adsabs.harvard.edu/abs/2021MNRAS.505.1495L} {505, 1495}

\bibitem[\protect\citeauthoryear{{Lee} \& {Baraffe}}{{Lee} \&
  {Baraffe}}{1995}]{1995A&A...301..419L}
{Lee} U.,  {Baraffe} I.,  1995, \aap, \href
  {https://ui.adsabs.harvard.edu/abs/1995A&A...301..419L} {301, 419}

\bibitem[\protect\citeauthoryear{{Lee} \& {Saio}}{{Lee} \&
  {Saio}}{2020}]{2020MNRAS.497.4117L}
{Lee} U.,  {Saio} H.,  2020, \mn@doi [\mnras] {10.1093/mnras/staa2250}, \href
  {https://ui.adsabs.harvard.edu/abs/2020MNRAS.497.4117L} {497, 4117}

\bibitem[\protect\citeauthoryear{{Lightkurve Collaboration}
  et~al.,}{{Lightkurve Collaboration} et~al.}{2018}]{2018ascl.soft12013L}
{Lightkurve Collaboration} et~al., 2018, {Lightkurve: Kepler and TESS time
  series analysis in Python}, Astrophysics Source Code Library (\mn@eprint
  {ascl} {1812.013})

\bibitem[\protect\citeauthoryear{{Moe} \& {Di Stefano}}{{Moe} \& {Di
  Stefano}}{2017}]{2017ApJS..230...15M}
{Moe} M.,  {Di Stefano} R.,  2017, \mn@doi [\apjs] {10.3847/1538-4365/aa6fb6},
  \href {https://ui.adsabs.harvard.edu/abs/2017ApJS..230...15M} {230, 15}

\bibitem[\protect\citeauthoryear{{Moon} \& {Dworetsky}}{{Moon} \&
  {Dworetsky}}{1985}]{1985MNRAS.217..305M}
{Moon} T.~T.,  {Dworetsky} M.~M.,  1985, \mn@doi [\mnras]
  {10.1093/mnras/217.2.305}, \href
  {https://ui.adsabs.harvard.edu/abs/1985MNRAS.217..305M} {217, 305}

\bibitem[\protect\citeauthoryear{{Mowlavi}, {Barblan}, {Saesen}  \&
  {Eyer}}{{Mowlavi} et~al.}{2013}]{2013A&A...554A.108M}
{Mowlavi} N.,  {Barblan} F.,  {Saesen} S.,   {Eyer} L.,  2013, \mn@doi [\aap]
  {10.1051/0004-6361/201321065}, \href
  {https://ui.adsabs.harvard.edu/abs/2013A&A...554A.108M} {554, A108}

\bibitem[\protect\citeauthoryear{{Murphy}, {Moe}, {Kurtz}, {Bedding},
  {Shibahashi}  \& {Boffin}}{{Murphy} et~al.}{2018}]{2018MNRAS.474.4322M}
{Murphy} S.~J.,  {Moe} M.,  {Kurtz} D.~W.,  {Bedding} T.~R.,  {Shibahashi} H.,
   {Boffin} H. M.~J.,  2018, \mn@doi [\mnras] {10.1093/mnras/stx3049}, \href
  {https://ui.adsabs.harvard.edu/abs/2018MNRAS.474.4322M} {474, 4322}

\bibitem[\protect\citeauthoryear{{Murphy}, {Hey}, {Van Reeth}  \&
  {Bedding}}{{Murphy} et~al.}{2019}]{2019MNRAS.485.2380M}
{Murphy} S.~J.,  {Hey} D.,  {Van Reeth} T.,   {Bedding} T.~R.,  2019, \mn@doi
  [\mnras] {10.1093/mnras/stz590}, \href
  {https://ui.adsabs.harvard.edu/abs/2019MNRAS.485.2380M} {485, 2380}

\bibitem[\protect\citeauthoryear{{Neiner}, {de Batz}, {Cochard}, {Floquet},
  {Mekkas}  \& {Desnoux}}{{Neiner} et~al.}{2011}]{Neiner2011}
{Neiner} C.,  {de Batz} B.,  {Cochard} F.,  {Floquet} M.,  {Mekkas} A.,
  {Desnoux} V.,  2011, \mn@doi [\aj] {10.1088/0004-6256/142/5/149}, \href
  {http://adsabs.harvard.edu/abs/2011AJ....142..149N} {142, 149}

\bibitem[\protect\citeauthoryear{{Paxton}, {Bildsten}, {Dotter}, {Herwig},
  {Lesaffre}  \& {Timmes}}{{Paxton} et~al.}{2011}]{pax11}
{Paxton} B.,  {Bildsten} L.,  {Dotter} A.,  {Herwig} F.,  {Lesaffre} P.,
  {Timmes} F.,  2011, \mn@doi [\apjs] {10.1088/0067-0049/192/1/3}, \href
  {https://ui.adsabs.harvard.edu/abs/2011ApJS..192....3P} {192, 3}

\bibitem[\protect\citeauthoryear{{Paxton} et~al.,}{{Paxton}
  et~al.}{2013}]{pax13}
{Paxton} B.,  et~al., 2013, \mn@doi [\apjs] {10.1088/0067-0049/208/1/4}, \href
  {http://adsabs.harvard.edu/abs/2013ApJS..208....4P} {208, 4}

\bibitem[\protect\citeauthoryear{{Paxton} et~al.,}{{Paxton}
  et~al.}{2015}]{pax15}
{Paxton} B.,  et~al., 2015, \mn@doi [\apjs] {10.1088/0067-0049/220/1/15}, \href
  {https://ui.adsabs.harvard.edu/abs/2015ApJS..220...15P} {220, 15}

\bibitem[\protect\citeauthoryear{{Pecaut} \& {Mamajek}}{{Pecaut} \&
  {Mamajek}}{2013}]{2013ApJS..208....9P}
{Pecaut} M.~J.,  {Mamajek} E.~E.,  2013, \mn@doi [\apjs]
  {10.1088/0067-0049/208/1/9}, \href
  {https://ui.adsabs.harvard.edu/abs/2013ApJS..208....9P} {208, 9}

\bibitem[\protect\citeauthoryear{{Richardson} et~al.,}{{Richardson}
  et~al.}{2021}]{2021MNRAS.508.2002R}
{Richardson} N.~D.,  et~al., 2021, \mn@doi [\mnras] {10.1093/mnras/stab2759},
  \href {https://ui.adsabs.harvard.edu/abs/2021MNRAS.508.2002R} {508, 2002}

\bibitem[\protect\citeauthoryear{{Rivinius}, {Baade}  \& {Carciofi}}{{Rivinius}
  et~al.}{2016}]{Rivinius2016}
{Rivinius} T.,  {Baade} D.,   {Carciofi} A.~C.,  2016, \mn@doi [\aap]
  {10.1051/0004-6361/201628411}, \href
  {https://ui.adsabs.harvard.edu/abs/2016A&A...593A.106R} {593, A106}

\bibitem[\protect\citeauthoryear{{Saio}, {Kurtz}, {Murphy}, {Antoci}  \&
  {Lee}}{{Saio} et~al.}{2018}]{2018MNRAS.474.2774S}
{Saio} H.,  {Kurtz} D.~W.,  {Murphy} S.~J.,  {Antoci} V.~L.,   {Lee} U.,  2018,
  \mn@doi [\mnras] {10.1093/mnras/stx2962}, \href
  {https://ui.adsabs.harvard.edu/abs/2018MNRAS.474.2774S} {474, 2774}

\bibitem[\protect\citeauthoryear{{Salmon}, {Montalb{\'a}n}, {Reese}, {Dupret}
  \& {Eggenberger}}{{Salmon} et~al.}{2014}]{2014A&A...569A..18S}
{Salmon} S.~J.~A.~J.,  {Montalb{\'a}n} J.,  {Reese} D.~R.,  {Dupret} M.~A.,
  {Eggenberger} P.,  2014, \mn@doi [\aap] {10.1051/0004-6361/201323259}, \href
  {https://ui.adsabs.harvard.edu/abs/2014A&A...569A..18S} {569, A18}

\bibitem[\protect\citeauthoryear{{Semaan}, {Hubert}, {Zorec},
  {Guti{\'e}rrez-Soto}, {Fr{\'e}mat}, {Martayan}, {Fabregat}  \&
  {Eggenberger}}{{Semaan} et~al.}{2018}]{Semaan2018}
{Semaan} T.,  {Hubert} A.~M.,  {Zorec} J.,  {Guti{\'e}rrez-Soto} J.,
  {Fr{\'e}mat} Y.,  {Martayan} C.,  {Fabregat} J.,   {Eggenberger} P.,  2018,
  \mn@doi [\aap] {10.1051/0004-6361/201629243}, \href
  {https://ui.adsabs.harvard.edu/abs/2018A&A...613A..70S} {613, A70}

\bibitem[\protect\citeauthoryear{{Sullivan} et~al.,}{{Sullivan}
  et~al.}{2015}]{2015ApJ...809...77S}
{Sullivan} P.~W.,  et~al., 2015, \mn@doi [\apj] {10.1088/0004-637X/809/1/77},
  \href {https://ui.adsabs.harvard.edu/abs/2015ApJ...809...77S} {809, 77}

\bibitem[\protect\citeauthoryear{{Szewczuk} \&
  {Daszy{\'n}ska-Daszkiewicz}}{{Szewczuk} \&
  {Daszy{\'n}ska-Daszkiewicz}}{2017}]{2017MNRAS.469...13S}
{Szewczuk} W.,  {Daszy{\'n}ska-Daszkiewicz} J.,  2017, \mn@doi [\mnras]
  {10.1093/mnras/stx738}, \href
  {https://ui.adsabs.harvard.edu/abs/2017MNRAS.469...13S} {469, 13}

\bibitem[\protect\citeauthoryear{{Van Reeth}, {Tkachenko}  \& {Aerts}}{{Van
  Reeth} et~al.}{2016}]{2016A&A...593A.120V}
{Van Reeth} T.,  {Tkachenko} A.,   {Aerts} C.,  2016, \mn@doi [\aap]
  {10.1051/0004-6361/201628616}, \href
  {https://ui.adsabs.harvard.edu/abs/2016A&A...593A.120V} {593, A120}

\bibitem[\protect\citeauthoryear{{Vieira}, {Carciofi}, {Bjorkman}, {Rivinius},
  {Baade}  \& {R{\'\i}mulo}}{{Vieira} et~al.}{2017}]{Vieira2017}
{Vieira} R.~G.,  {Carciofi} A.~C.,  {Bjorkman} J.~E.,  {Rivinius} T.,  {Baade}
  D.,   {R{\'\i}mulo} L.~R.,  2017, \mn@doi [\mnras] {10.1093/mnras/stw2542},
  \href {https://ui.adsabs.harvard.edu/abs/2017MNRAS.464.3071V} {464, 3071}

\makeatother
\end{thebibliography}

\bsp	
\label{lastpage}
\end{document}